\documentclass[
    final            
  ,numberedheadings 
  ]
  {aipproc}

\layoutstyle{6x9}

\begin{document}

\title{AGN outflow feedback: Constraints from variability}

\author{R.G. Detmers}{
  address={SRON Netherlands Institute for Space Research, Sorbonnelaan 2, 3584 CA, The Netherlands}}
 \author{J.S. Kaastra}{
   address={SRON Netherlands Institute for Space Research, Sorbonnelaan 2, 3584 CA, The Netherlands}}

\begin{abstract}
We present an overview on how variability can be used to constrain the location of the ionized outflow in nearby Active Galactic Nuclei using high-resolution X-ray spectroscopy. Without these constraints on the location of the outflow, the kinetic luminosity and mass loss rate can not be determined. We focus on the Seyfert 1 galaxy NGC 5548, which is arguably the best studied AGN on a timescale of 10 years. Our results show that frequent observations combined with long term monitoring, such as with the \textit{Rossi X-ray Timing Explorer (RXTE)} satellite, are crucial to investigate the effects of these outflows on their surroundings.  
\end{abstract}

\maketitle


\section{Introduction}

Active Galactic Nuclei (AGN) outflows are thought to play an important role in feedback processes. The growth of the supermassive black hole is connected to the growth of the bulge and the interstellar and intergalactic medium are thought to be enriched by these AGN outflows \cite{DiMatteo05}. However the main problem is that we do not know the location or origin of these outflows. The kinetic luminosity and mass loss rate can not be accurately determined if the location is unknown. By studying these outflows in the X-ray regime with high-resolution grating spectrometers over multiple years, we can constrain the location of the outflow by using the intrinsic variability of the source \cite{Netzer03, Nicastro07}. We will focus on one particular Seyfert 1 galaxy, NGC 5548, because it is the best studied AGN, with high-resolution X-ray data spanning almost 8 years in total \cite{Detmers08}.

\section{Observations}

\begin{figure}
  \label{fig:rxte}
  \includegraphics[width=8cm]{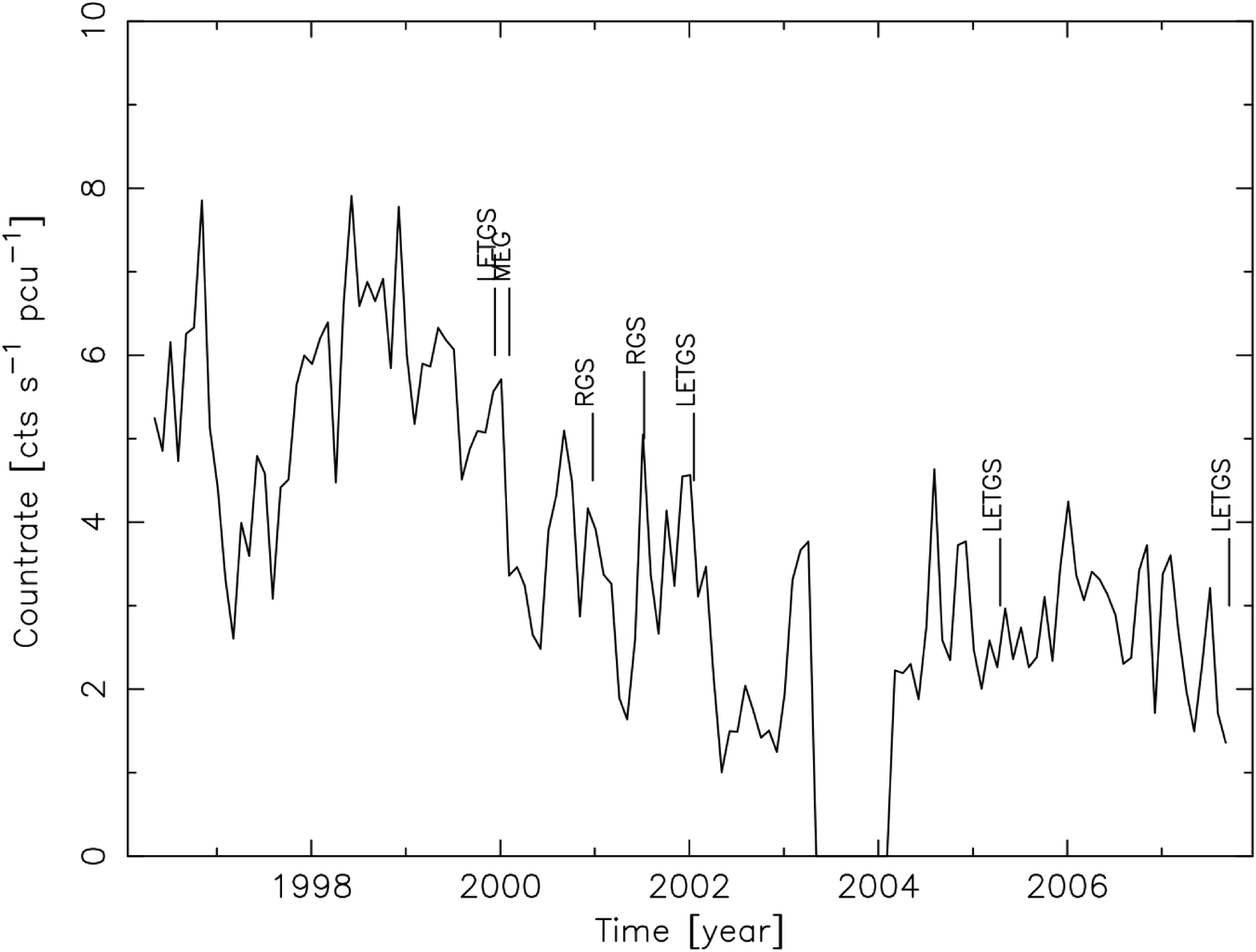}
  \caption{The \textit{RXTE} 2$-$10 keV lightcurve from 1996 to 2007. The data have been binned to one month intervals and the different high-resolution spectral observations are indicated as well. There is a small data gap between Apr 2003 and Jan 2004.}
\end{figure}

We use observations made by the \textit{Chandra}$-$LETGS and XMM$-$\textit{Newton} RGS to analyze the X-ray spectrum of NGC 5548. In order to constrain the continuum history of the source we have also used \textit{RXTE} data to accurately track the variations of the continuum flux. Fig. \ref{fig:rxte} shows the \textit{RXTE} lightcurve from 1996 to 2007. All high-resolution X-ray observations are also indicated. We focus here on the 2002 and 2005 observations, since a large drop in flux occurred between these observations. 

\section{Spectral analysis and method}

After 2002 the source experienced a large drop in flux and although variations are still observed, the average flux level after 2002 is a factor of 2$-$3 lower than before.
This decrease in flux will have an effect on the warm absorber gas in our line of sight. We expect the gas to start recombining to a lower ionized state as a result, but only after a delay time. 
The delay, or recombination time of the gas $\tau_{\mathrm{rec}}$ is determined by the density $n$ of the gas and the recombination rate $\alpha_{r}$ of a certain ion in the following way:

\begin{equation}		\label{trec}
\centering
      \tau_{\mathrm{rec}}(X_{i}) = \left({\alpha_{\mathrm{r}}(X_{i})n \left[\frac{f(X_{i+1})}{f(X_{i})} - \frac{\alpha_{\mathrm{r}}(X_{i-1})}{\alpha_{\mathrm{r}}(X_{i})}\right]}\right)^{-1}, 
\end{equation}
where $\alpha_{\mathrm{r}}(X_{i})$ is the recombination rate from ion $X_{i=1}$ to ion $X_{i}$ and $f(X_{i})$ is the fraction of element $X$ in ionization state $i$. 
If we know $\tau_{\mathrm{rec}}$, then we also know the density $n$. This density estimate can then be used together with the ionizing luminosity $L_{ion}$ and ionization state $\xi$ to determine the distance of the absorbing gas:

\begin{equation}		\label{xi}
\centering
      \xi = \frac{L_{ion}}{n R^{2}}. 
\end{equation}
Here $L_{\mathrm{ion}}$ is the 1 $-$ 1000 Rydberg luminosity, $n$ is the density of the gas and $R$ is the distance from the ionizing source. In practice however, the recombination time can not be exactly determined and only an upper limit can be obtained. This results in a lower limit on the density $n$ and therefore in an upper limit to the distance $R$.

\subsection{Photoionized outflow}

\begin{figure}[tbp]
    \includegraphics[width=4cm,angle=-90]{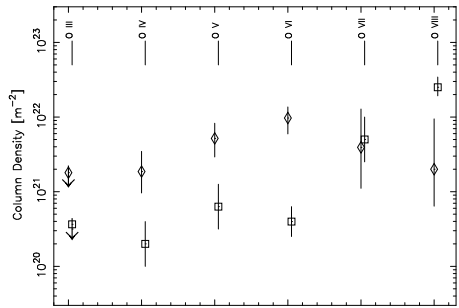}
    \caption{\label{fig:column}
         Total oxygen column densities as measured with the LETGS instrument in 2002 (rectangles) and 2005 (diamonds). 1$\sigma$ errors are shown as well, apart from ions for which an upper limit to the total column density can be measured (arrows).}
\end{figure}

The photoionized outflow consists of multiple components spanning a wide range in ionization states. We focus here on the oxygen ions, since those ions produce the strongest and most prominent lines in this source. By comparing the observed column densities between the two observations, we can detect changes in the warm absorber and investigate whether they are consistent with recombination. Fig. \ref{fig:column} shows the observed column densities for the detected oxygen ions. It can be seen that the highly ionized O\,VIII has decreased in column density, while the lower ionized O\,IV$-$O\,VI have increased in column density. This is a clear sign of recombination and therefore we can put a constraint on $\tau_{\mathrm{rec}}$. The time between the observations is 1160 days, so the (conservative) upper limit to $\tau_{\mathrm{rec}}$ is also 1160 days. This results in an upper limit to the distance of 6 pc for the gas which produces the O\,VIII absorption. The upper limit to the distance for the other ions is larger than 6 pc, so O\,VIII yields tightest constraint on the location of the warm absorber.

\subsection{Feedback estimates}

With a constraint on the distance to the warm absorber, we can estimate the mass outflow rate. The column density of the gas which is responsible for the O\,VIII absorption is 10$^{25}$ m$^{-2}$ and the lower limit to the density is 4 $\times$ 10$^{9}$ m$^{-3}$. If we assume that the warm absorber consists of spherical clouds, then the mass loss rate is: 

\begin{equation}		\label{mdot}
\centering
      \frac{dM_{out}}{dt} \sim \pi(\Delta\,R)^{2}\,v\,m_{p}\,n, 
\end{equation} 
with $\Delta\,R$ = 0.5 $\frac{N_{H}}{n}$, $m_{p}$ the proton mass and $v$ the outflow velocity (1000 km s$^{-1}$ for this component). The mass outflow rate we obtain in this way is 10$^{-5}$ M$_{\odot}$\,yr$^{-1}$, which is a lower limit. The solid angle $\Omega$ of the outflow is just $\pi\,\left(\frac{\Delta\,R}{R}\right)^{2}$. Again using the above limits on $\Delta\,R$ and $R$, we obtain an $\Omega$ of 10$^{-5}$. 
The same exercise can be done for the other components as well, but the O\,VIII component places the tightest constraints on the warm absorber properties.
  
\section{Discussion}

There are two main problems which prevent us from obtaining accurate values for the warm absorber properties (mass outflow, kinetic luminosity and solid angle) at the moment. The first is the recombination timescale, for which we only have upper limits so far. The second is the geometry of the outflow, which is not known. For example we have derived a solid angle of 10$^{-5}$ assuming an outflow which consists of spherical blobs of gas. However about 50 $\%$ of the Seyfert 1 galaxies show a warm absorber. This is inconsistent with the picture of such a small solid angle, unless the outflow bends and is in the form of narrow streams of gas. 
Using the current data which are available in the archives for both high-resolution X-ray observations and high-resolution UV data and combining them with time-dependent photoionization models, we can try to obtain a more coherent picture of the feedback processes in local Seyfert galaxies. 

\section{Conclusions}
Long-term monitoring of AGN in combination with high-resolution X-ray spectroscopy is key to constrain the location of the warm absorber outflows. Once the location has been determined, the feedback strength of the outflow (mass outflow rate, kinetic luminosity) can not be accurately determined without knowing the geometry of the outflow. General estimates can be made, either by assuming that the mass outflow rate is less than the accretion rate or by assuming a specific geometry for the outflow, but the key to unravelling the importance of feedback in these local AGN is variability, combined with multi-wavelength observations to constrain the geometry of these outflows. Only then can a full picture of feedback in local AGN be obtained.

\begin{theacknowledgments}
 RGD wishes to thank Elisa Costantini for useful discussions regarding the paper. SRON is supported financially by NWO, The Netherlands Organization for Scientific Research.
\end{theacknowledgments}


\bibliography{bibfiles}

\begin{thebibliography}{4}
\expandafter\ifx\csname natexlab\endcsname\relax\def\natexlab#1{#1}\fi
\providecommand{\enquote}[1]{``#1''}
\expandafter\ifx\csname url\endcsname\relax
  \def\url#1{\texttt{#1}}\fi
\expandafter\ifx\csname urlprefix\endcsname\relax\def\urlprefix{URL }\fi
\providecommand{\eprint}[2][]{\url{#2}}

\bibitem[{Di Matteo} et~al.(2005)]{DiMatteo05}
T.~{Di Matteo}, V.~{Springel}, and L.~{Hernquist}, \emph{Nature} \textbf{433},
  604--607 (2005).

\bibitem[{Netzer} et~al.(2003)]{Netzer03}
H.~{Netzer}, S.~{Kaspi}, E.~{Behar}, W.~N. {Brandt}, D.~{Chelouche}, I.~M.
  {George}, D.~M. {Crenshaw}, J.~R. {Gabel}, F.~W. {Hamann}, S.~B. {Kraemer},
  G.~A. {Kriss}, K.~{Nandra}, B.~M. {Peterson}, J.~C. {Shields}, and T.~J.
  {Turner}, \emph{ApJ} \textbf{599}, 933--948 (2003).

\bibitem[{Nicastro} et~al.(2007)]{Nicastro07}
F.~{Nicastro}, M.~{Elvis}, N.~{Brickhouse}, Y.~{Krongold}, L.~{Binette}, and
  S.~{Mathur}, \enquote{{Time-evolving Photoionization: The Thin and Compact
  X-ray Wind of NGC 4051},} in \emph{The Central Engine of Active Galactic
  Nuclei}, edited by L.~C. {Ho}, and J.-W. {Wang}, 2007, vol. 373 of
  \emph{ASPC}, p. 301.

\bibitem[{Detmers} et~al.(2008)]{Detmers08}
R.~G. {Detmers}, J.~S. {Kaastra}, E.~{Costantini}, I.~M. {McHardy}, and
  F.~{Verbunt}, \emph{A\&A} \textbf{488}, 67--72 (2008).

\end{thebibliography}
\bibliographystyle{aipproc}   


\end{document}